\documentclass[pre,twocolumn,aps,superscriptaddress,showpacs,floatfix]{revtex4}
\usepackage{amsmath}
\usepackage{graphicx}
\numberwithin{equation}{section}
\begin{document}
\title{Recycled Noise Rectification: A Dumb Maxwell's Daemon}
\author{M. Borromeo}
\affiliation{Dipartimento di Fisica, Universit\`a di Perugia,
I-06123 Perugia, Italy}
\affiliation{Istituto Nazionale di Fisica Nucleare, Sezione di Perugia,
I-06123 Perugia, Italy}
\author{S. Giusepponi and F. Marchesoni}
\affiliation{Dipartimento di Fisica, Universit\`a di Camerino,
I-62032 Camerino, Italy}
\date{\today}
\begin{abstract}
The one dimensional motion of a massless Brownian particle on a symmetric periodic substrate can be rectified by re-injecting its driving noise through a realistic recycling procedure. If the recycled noise is multiplicatively coupled to the substrate, the ensuing feed-back system works like a passive Maxwell's daemon, capable of inducing a net current that depends on both the delay and the autocorrelation times of the noise signals. Extensive numerical simulations show that the underlying rectification mechanism is a resonant nonlinear effect: The observed currents can be optimized for an appropriate choice of the recycling parameters with immediate application to the design of nanodevices for particle transport.
\end{abstract}
\maketitle

\section{Introduction}

When control or spurious signals, either periodic or noisy, are injected into an extended system, they can couple a variable of interest $x(t)$ through different channels, additively or multiplicatively, alike. While being transmitted across the system components, an input signal may undergo \cite{risken,papoulis}:

{\it (a) time delay}, due to the combination of diverse propagation or transduction mechanisms. As a result, an input signal $\eta(t)$ can split into two or more forcing signals $\xi_i(t)$ acting on $x(t)$, each with a time delay $\tau_d^{(i)}$, i.e.
\begin{equation}
\label{1.1}
\xi_i(t)=\sqrt{Q_i}\eta(t- \tau_d^{(i)}).
\end{equation}

\begin{figure}[ht]
\includegraphics*[width=7.0cm]{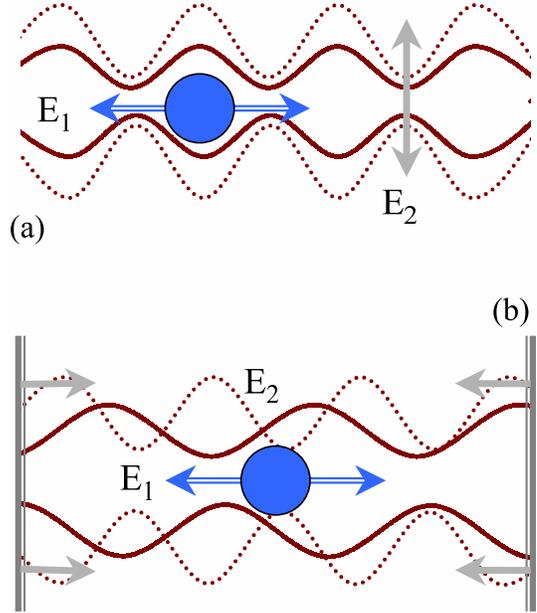}
\caption{\label{F1} (Color online) Effective action of an external field of force
on a Brownian particle in a narrow channel: (a) The external field splits into a direct, $E_1(t)$, and a transmitted component, $E_2(t)$, that couple to the particle with relative delay $\tau_c$; (b) The field couples to the particle both directly, $E_1(t)$, and modifying its environment through the local field $E_2(t)$. In both cases $E_2(t)$ modulates the channel geometry (multiplicative coupling).}
\end{figure}

Let us consider, for instance, an electric signal acting on a particle of coordinate $x(t)$ in a narrow channel after propagating e.g. through two different nondispersive media, as illustrated in Fig. \ref{F1}(a). Regardless of the nature of the input signal, a wave train or random pulses, the time delay $\tau_d$ of $E_2(t)$ relative to $E_1(t)$ depends of the electromechanical properties of the medium surrounding the channel; moreover, the two drives are likely to couple the particle differently: $E_1(t)$ pulls the particle horizontally and, therefore, additively, while $E_2(t)$ propagates normally to the channel, thus modulating multiplicatively its effective substrate potential;

{\it (b) time correlation}, due to the finite response time of the system components. The relevant response, or transfer functions modulate the amplitude of the signals $\xi_i(t)$ effectively coupled to the variable of interest $x(t)$ and, more importantly, determine a finite autocorrelation/decay time $\tau_c^{(i)}$ for each $\xi_i(t)$, depending on its coupling channel \cite{risken,JSM}.

Figure \ref{F1}(b) shows the ideal case of a particle diffusing through a one dimensional (1D) pore, or channel in a homogeneous dielectric medium with relaxation time $\tau_c$; the sample is placed between the plates of a capacitor subjected to a variable voltage. The particle is thus directly coupled to the time-dependent electric field $E_1(t)$; however, its mobility in the $x$ direction changes in time as the channel cross-section is modulated by the electrostriction effects induced by the capacitor. The potential energy of the particle along its path is thus a function of the local field $E_2(t)$, which, in turn, can be regarded as a dielectric response to the input $E_1(t)$. As long as the frequency dependence of $\tau_c$ can be neglected (low frequency regime), amplitude and phase-shift of each spectral component of $E_2(t)$ can be explictly computed in linear response theory \cite{risken}. If $E_1(t)$ is a random signal with correlation time $\tau_c^{(1)}$, then $E_2(t)$ is an autocorrelated noise with time constant $\tau_c^{(2)}=\tau_c+\tau_c^{(1)}$ \cite{JSM}, while the primary, $E_1(t)$, and the secondary signal, $E_2(t)$, are crosscorrelated with effective time constant $\tau_c$. If $E_1(t)$ is an ac field with angular frequency $\omega$, the  signal $E_2(t)$ modulating the channel develops a time delay $\tau_c \simeq \arctan(\omega\tau_c)/\omega$ \cite{SR}.

In this paper we work with stationary Gaussian noise sources and restrict ourselves to the case of two distinct coupling channels, only, so that a given noise source $\eta(t)$ splits into two colored driving noises, $\eta(t) \to \xi_1(t), \xi_2(t)$, with correlation times $\tau_c^{(1)}$ and $\tau_c^{(2)}$, and delay times $\tau_d^{(1)}$ and $\tau_d^{(2)}$. To simplify our notation we further assume that: (i) 
$\tau_c^{(1)}=\tau_c^{(2)}=\tau_c \geq 0$; (ii) $\tau_d^{(2)}-\tau_d^{(1)}=\tau_d \geq 0$, which for a stationary $\eta(t)$ is equivalent to setting $\tau_d^{(1)}=0$ and $\tau_d^{(2)}=\tau_d$. 

For the purpose of numerical simulation, conditions (i) and (ii) are satisfied by noises of the form (\ref{1.1}) with $\eta(t)$ a Gaussian noise with autocorrelation function \cite{JSM}
\begin{equation}
\label{1.2}
\langle \eta(t)\eta(0)\rangle = e^{-|t|/\tau_c}/\tau_c.
\end{equation}
All correlation functions $\langle \xi_i(t)\xi_j(0)\rangle$, with $i,j=1,2$, can be readily expressed in terms of the autocorrelation function (\ref{1.2}). According to this notation, for practical purposes we term $\xi_1(t)$ the primary or source noise, and $\xi_2(t)$ the secondary or recycled noise [with no reference to $\eta(t)$].

Physical systems driven by {\it delayed correlated} noises are rather common in nature, typical examples being the propagation of charge density waves \cite{CDW,HM}, the migration of both pointlike and linear defects in crystalline materials \cite{defect}, the transport of nano-particles in biological \cite{karger} and artificial channels \cite{pores}, the manipulation of vortex lines in superconducting devices \cite{tonomura} and colloidal particles along 1D tracks \cite{bechinger}, to mention but a few.

In most models discussed in the literature, the input signal $\eta(t)$ is time periodic and so are the two (or more) driving terms $\xi_i(t)$ \cite{HM,riken_H,russi}: A given phase-lag between two additive signals \cite{HM,riken_H} or between an additive and a multiplicative signal \cite{sergey,russi} may breach the spatial symmetry of the underlying $x$ dynamics, thus inducing a net current $\langle \dot x \rangle$. [Note that the average of $\langle \dot x \rangle$ over a uniform distribution of the phase-lag vanishes (spontaneous symmetry breaking mechanism).]
Here, we consider similar models with the difference that the ac drives are replaced by the noises $\xi_i(t)$ introduced above. 

Our models with two (or more) delayed noises should not be mistaken for the nonlinear systems with delay recently investigated by several groups in the context of stochastic resonance \cite{pikovsky,masoller}: These authors proposed to control the dynamics of $x(t)$ by utilizing appropriate functions of $x(t-\tau_d)$ as feedback terms. This is a well-established control technique in physics and electrical engineering \cite{papoulis}. Extended experimental apparatuses, where both types of delays (from recycling and feedback) must be taken into account, are, for instance, the gravitational-wave interferometers, like the VIRGO detector \cite{virgo}. Here, an external signal $\eta(t)$, e.g., a seismic disturbance, enters the antenna by creeping through its mechanical suspensions and dampers and eventually combines with the intrinsic electronic and photonic noises of the apparatus, so that the detection signal $x(t)$, corresponding to the mirror displacement induced by the gravitational signal, is additively and multiplicatively affected by $\eta(t)$ at different times. Moreover, the control loops that maintain the alignment and the locking of the interferometer, make use of sophisticated feedback techniques also involving $x(t)$.

In all examples mentioned so far, the multiple action of an external disturbance is often regarded as a nuisance;  on modeling the system of interest one simply takes notice of its presence and tries to minimize its impact on the system response. In our approach we took a more ''pro-active'' stance, namely, we propose to tap the primary noise source $\xi_1(t)$  and re-inject the recycled noise signal $\xi_2(t)$ into the system so as to control its response. At variance with an earlier attempt \cite{confinement}, we consider here the motion of a massless Brownian particle diffusing on a symmetric 1D periodic substrate; for appropriate choices of the time constants $\tau_c$ and $\tau_d$ the particle drifts sidewise with net velocity $\langle \dot x \rangle$. Such a mechanism can be viewed as an automated Maxwell's daemon, where the rectification of the primary noise signal $\xi_1$ is achieved through a pre-assigned recycling protocol: Lacking the dexterity originally assumed by Maxwell, our ''dumb'' stochastic daemon performs its chore ''in average'' and with reduced (but dependable) efficiency. Note that rectification of a primary noise by nonlinear interference with a recycled image of itself can hardly be assimilated to a ratchet mechanism \cite{ratchet}, as here no spatial substrate asymmetry is required.

This paper is organized as follows: In Sec. II the primary, $\xi_1$, 
and the secondary noise, $\xi_2$, are coupled additively to an overdamped variable $x$ bound to either a parabolic well (Ornstein-Uhlenbeck process) or a quartic double well (Kramers problem) \cite{risken}. In Sec. III we simulate numerically a massless Brownian particle diffusing, under the action of an additive primary noise $\xi_1$, in a 1D cosine potential modulated in amplitude by the recycled noisy term $\xi_2$. The particle rectification current $\langle \dot x \rangle$ attains an optimal intensity in an appropriate range of the time constants $\tau_c$ and $\tau_d$. In Sec. IV we discuss the dependence of $\langle \dot x \rangle$ on the remaining simulation parameters in view of the potential applications of our rectification scheme to the design of nano-devices for particle transport.

\section{Noise Recycling}

The technique of noise recycling is well-known in laser interferometry \cite{virgo}: One extracts
a signal from the fluctuation source to be suppressed and re-injects it into the system after appropriate manipulation; the process involves a certain number of steps, like analogue or digital acquisition, usage of filters and actuators, etc, which eventually cause a delay of the feedback $\xi_2(t)$ with respect to the primary signal $\xi_1(t)$.

In order to familiarize the reader with the interplay of two delayed noises, we first consider the Ornstein-Uhlenbeck process
\begin{equation}
\label{2.1}
\dot x = -a x +\xi_1(t) \pm \xi_2(t),
\end{equation}
where $\xi_1(t)=\sqrt{Q_1} \eta(t)$ and $\xi_2(t)=\sqrt{Q_2} \eta(t-\tau_d)$, with $\tau_d \geq 0$, and
$\eta(t)$ denotes a zero-mean, white Gaussian noise with autocorrelation function $\langle \eta(t) \eta(0) \rangle = 2 \delta (t)$. The stochastic differential equation (SDE) (\ref{2.1}) can be treated analytically by mens of standard techniques \cite{risken,papoulis}. In particular,
\begin{eqnarray}
\label{2.2}
\langle [x(t) &-& x(0)]^2\rangle = \frac{Q_1+Q_2}{a}\, (1-e^{-2at}) \\
&\pm& 2\frac{\sqrt{Q_1 Q_2}}{a}\, e^{-a\tau_d} [1-e^{-2a(t-\tau_d)}]\Theta(t-\tau_d), \nonumber
\end{eqnarray}
where $\Theta(x)$ is the Heaviside step-function and asymptotically
\begin{eqnarray}
\label{2.3}
\langle x^2 \rangle&\equiv& \lim_{t\to \infty} \langle [x(t) - x(0)]^2\rangle \\ \nonumber
&=& \frac{1}{a}[(Q_1 + Q_2) \pm 2\sqrt{Q_1 Q_2}e^{-a\tau_d}]. 
\end{eqnarray}
Analogously, one can compute the autocorrelation function
\begin{eqnarray}
\label{2.4}
& &C(t) = \lim_{\tau \to \infty} \langle x(t+\tau) x(\tau) \rangle \\
&=& \frac{\sqrt{Q_1Q_2}}{a}\, e^{-at}\left [\frac{Q_1 + Q_2}{\sqrt{Q_1 Q_2}} \pm \left (e^{-a\tau_d}+
e^{-a(|t-\tau_d| - t)} \right ) \right ] \nonumber
\end{eqnarray}
with $\tau_d \geq 0$. Note that $C(-t)=C(t)$ and $C(0)=\langle x^2 \rangle \geq 0$, as it should.

\begin{figure}[ht]
\includegraphics*[width=8.0cm]{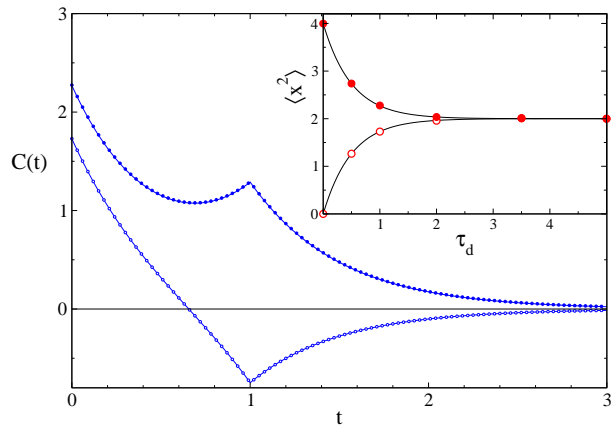}
\caption{\label{F2} (Color online) Autocorrelation function $C(t)$ of the process (\ref{2.1}) for
$Q_1=Q_2=2$ and $a=2$; the recycled noise $\xi_2$ is added to (filled circles) or subtracted (empty circles) from the primary noise $\xi_1$ after a delay time $\tau_d=1$. The solid curves represent the corresponding analytical predictions (\ref{2.4}). Inset: $C(0) = \langle x^2 \rangle$ vs. $\tau_d$;
notation and the remaining simulation parameters are as in the main panel.}
\end{figure}

In Fig. \ref{F2} we compare numerical simulation results with our predictions (\ref{2.3}), inset, and 
(\ref{2.4}), main panel: in all cases the corresponding data sets are indistinguishable. Notice that sampling the $x$-autocorrelation only for $t>\tau_d$, thus overlooking the possibility that the process can be driven by two delayed noises, would lead to overestimating $\langle x^2 \rangle$,
\begin{equation}
\label{2.5}
\langle x^2 \rangle=\frac{1}{a}[(Q_1 + Q_2) \pm 2\sqrt{Q_1 Q_2}\cosh(a\tau_d)],
\end{equation}
with respect to the correct value (\ref{2.3}).
Generalizations of the process (\ref{2.1}) to account for colored noises, with exponential autocorrelation functions of the type in Eq. (\ref{1.2}), and/or inertial effects, with $\dot x$ replaced by $\ddot x+\gamma \dot x$, can also be treated analytically. Harmonic analysis \cite{papoulis} yields  straightforward, though cumbersome expressions for $C(t)$. 

More suggestive is the {\it nonlinear} case of a massless particle bound to a quadratic double well potential
$V(x) = -ax^2/a + bx^4/4$ \cite{JSM,SR},
\begin{equation}
\label{2.6}
\dot x=ax-bx^3 +\xi_1(t) \pm \xi_2(t),
\end{equation}
with $\xi_1$ and $\xi_2$ defined as above, and $a,b>0$. Subjected to the random kicks by the fluctuation sources, the overdamped Brownian particle hops between two bistable states centered at $\pm \sqrt{a/b}$ and separated by the potential barrier $\Delta V=a^2/4b$. According to Kramers' rate theory, the relevant hopping time \cite{risken,SR} in the low noise regime is
\begin{equation}
\label{2.7}
T_K^{\pm}(\tau_d)=\frac{\pi\sqrt{2}}{a}\, \exp{\left [\frac{\Delta V} {D_{\pm}(\tau_d)}\right ]},
\end{equation}
where $D_{\pm}(\tau_d)$ is the effective noise intensity of $\xi_1(t) \pm \xi_2(t)$, to be determined next.

\begin{figure}[ht]
\includegraphics*[width=8.0cm]{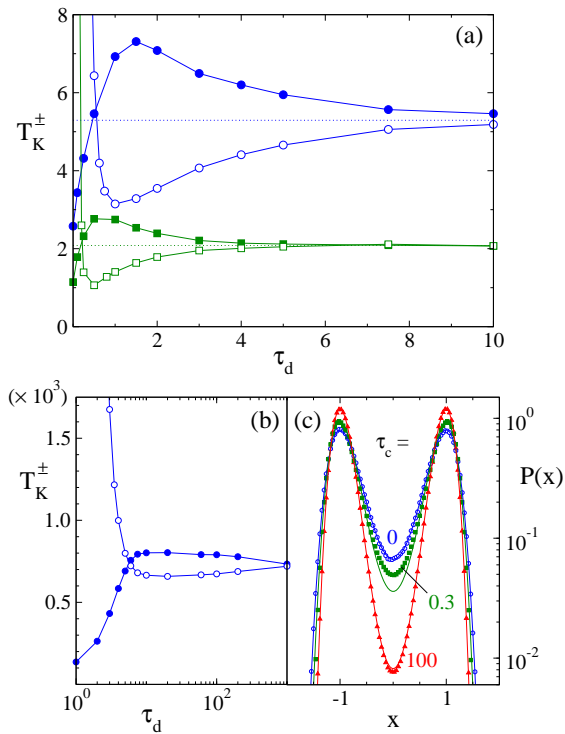}
\caption{\label{F3} (Color online) Activation process in the bistable potential
$V(x) = -ax^2/2 + bx^4/4$  driven by  $\xi_1(t) \pm \xi_2(t)$: (a), (b) $T_K^{+}$ (filled symbols) and  $T_K^{-}$  (empty symbols) vs. $\tau_d$ for $\tau_c=0$ and $Q_1=Q_2$; (c) Probability density $P(x)$ for  $\tau_c=0$,
$Q_1=Q_2=0.025$, and $\tau_d=0$ (circles), $0.3$ (squares), and $100$ (triangles).}
\end{figure}

For zero delay, $\tau_d=0$, $D_+= (\sqrt{Q_1}+\sqrt{Q_2})^2$ and $D_-=0$; for extended delays, $\tau_d \to \infty$, the two noises get completely uncorrelated, so that $D_{\pm}=Q_1 + Q_2$. As a consequence, the simulation results of Figs. \ref{F3}(a) and (b) reveal the following limiting behaviors of $T_K^{\pm}$: (i) For $\tau_d \to \infty$, both $T_K^{\pm}(\tau_d)$ tend to one limit, (\ref{2.7}), with
$D_{\pm}=Q_1 + Q_2$ -- perturbation corrections to $T_K^{\pm}(\infty)$ may be computed in powers of $D_{\pm}/ \Delta V$ \cite{risken}; (ii) On subtracting the recycled from the primary noise, the Kramers' time $T_K^{-}(\tau_d)$  tends to diverge; the effect gets amplified at small $\tau_d$, when the destructive interference of the two noises is the most effective; (iii) When adding $\xi_1$ and $\xi_2$ with $\tau_d \to 0$, the Kramers' formula (\ref{2.7}) still applies but for  $D_+= (\sqrt{Q_1}+\sqrt{Q_2})^2$. Of course, $T_K^{+}(0) < T_K^{\pm}(\infty)$.

The $\tau_d$ dependence of $T_K^{\pm}(\tau_d)$ for intermediate delays 
$\tau_d \sim \frac{1}{2}T_K^{\pm}(\infty)$ is more interesting: The curves $T_K^{+}(\tau_d)$ exhibit a broad peak, whereas the curves $T_K^{-}(\tau_d)$ develop a shallow dip  within approximately the same $\tau_d$ interval. This is a manifestation of the so-called resonant activation \cite{res.activation}. 
Introducing a delay $\tau_d$ when superimposing two noises $\xi_1$ and $\xi_2$ with the same sign ($+$ sign), hinders the hopping dynamics for $\tau_d \ll T_K^{\pm}(\infty)$; indeed, increasing $\tau_d$ decorrelates  $\xi_1$ and $\xi_2$ in time, i.e., diminishes the effective intensity  $D_+(\tau_d)$. This effect goes through a maximum for $\tau_d \sim \; \frac{1}{2}T_K^{+}(\tau_d)$: A strong kick of $\xi_1$ capable of making the particle reach the top of the barrier at $x=0$, will be counter-balanced in average by an opposite kick of $\xi_2$, which prevents the particle from falling into the other well. At even larger $\tau_d$, the action of the two noises grows totally uncorrelated and $T_K^{+}(\tau_d)$ approaches $T_K^{\pm}(\infty)$ from above. At variance, combining $\xi_1$ and $\xi_2$ with opposite signs ($-$ sign) produces a synchronization effect; for intermediate delays $\tau_d$ hopping occurs after shorter waiting times, hence the dip in the $T_K^{\pm}(\tau_d)$ curves.

On lowering $Q_1$ and $Q_2$ the Kramers' time increases exponentially; due to the prolonged sojourn at the well bottoms, memory effects in the particle dynamics are suppressed and so is the resonant behavior of $T_K^{\pm}(\tau_d)$. Finally, we stress that the $x$ probability densities $P(x)$ obey the Boltzmann law $P(x) \propto e^{-V(x)/D_{\pm}}$ only for $\tau_d \to 0$ and $\tau_d \to \infty$. For intermediate $\tau_d$ values the reduction of the dynamics (\ref{2.5}) to a 1D stationary process is questionable \cite{JSM}, especially in the vicinity of the barrier; in other words, fitting the simulation $P(x)$ curves of Fig. \ref{F3}(c) by means of the Boltzmann law does not rigorously define $D_{\pm}$ in Eq. (\ref{2.7}) -- but rather allows to extract a working estimate of it.

Note that laser systems with opto-electronic feedbacks, like those described in Refs. \cite{masoller,masoller2}, are already capable of experimentally investigating the Kramers' rate mechanism driven by delayed noises.

\section{A dumb Maxwell's daemon}

At variance with the models of Sec. II, the recycled noise $\xi_2$ can be re-injected so as to modulate the substrate on which the Brownian particle diffuses subjected to the primary noise $\xi_1$. In Ref. \cite{confinement} we investigated the confined process
\begin{equation}
\label{3.1}
\dot x=a[1+\xi_2(t)]x-bx^3 +\xi_1(t),
\end{equation}
with $\xi_1$ and $\xi_2$ defined as in Sec. II
and $V(x) = -ax^2/a + bx^4/4$ with $a,b>0$. The SDE (\ref{3.1}) is not symmetric under reflection $x \to -x$; this implies that, regardless of the delay $\tau_d$, the probability density $P(x)$ is asymmetric. By looking at the sign of the noises and the potential parameters, one concludes that for $\tau_d=0$ a positive drive $\xi_1$ (i.e., tilting the bistable potential $V(x)$ to the right) corresponds to rising the right-to-left barrier by a fraction proportional to $\xi_2=\xi_1$, and vice versa for $\xi_1 <0$. As a consequence the particle is more likely to get trapped in the left well and the negative peak of $P(x)$ overshoots the positive one ({\it gating mechanism} \cite{riken_H,confinement}).

Two conflicting effects determine the optimal $P(x)$ asymmetry: On one side, since the particle takes a finite Kramers' time to reach the top of the barrier, delaying $\xi_2$ with respect to $\xi_1$ synchronizes their activation effort thus improving the success rate of the noise assisted hopping; on the other side, increasing $\tau_d$ larger than $\tau_c$ makes the superposition of the two noises statistically incoherent and restores the spatial symmetry of $P(x)$. In conclusion, for low noise intensities, the density asymmetry is maximal around $\tau_d \sim \tau_c$ \cite{confinement}. 

In the following we investigate the Brownian motion on a sinusoidal substrate
\begin{equation}
\label{3.2}
\dot x=a[1+\xi_2(t)] \cos x +\xi_1(t) +\epsilon(t),
\end{equation}
where $\xi_1$ and $\xi_2$ are colored noises with correlation time $\tau_c$ and relative delay $\tau_d$,
and $\epsilon(t)$ is a Gaussian stationary noise with $\langle \epsilon (t)\rangle =0$, $\langle \epsilon (t) \eta(0)\rangle =0$, and $\langle \epsilon (t) \epsilon(0)\rangle =2D\delta(t)$. Throughout this section we switch off the thermal noise $\epsilon$, that is we set $D\equiv 0$.

\begin{figure}[ht]
\includegraphics*[width=8.0cm]{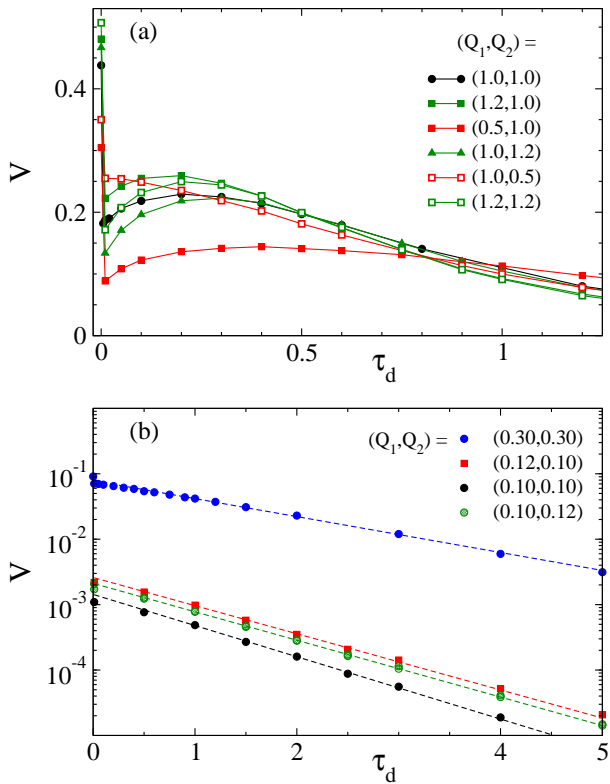}
\caption{\label{F4} (Color online) Characteristics curve $v(\tau_d) \equiv |\langle \dot x (\tau_d) \rangle|$ for the process (\ref{3.2}) with $a=1$, $D\equiv 0$, $\tau_c=0$, and different noise intensities $(Q_1,Q_2)$: (a) high noise levels, resonant tails; (b) low noise level, exponential tails versus fitting law (\ref{3.3}) (dashed lines). In both panels  $\langle \dot x (\tau_d) \rangle$ is negative.}
\end{figure}

By the same line of reasoning we outlined for the confined process (\ref{3.1}), we expect that the Brownian dynamics (\ref{3.2}) gets rectified to the left with negative net velocity $\langle \dot x (\tau_d) \rangle$. 
Numerical simulation confirms our expectations, i.e., $\langle \dot x (\tau_d) \rangle <0$. Figure \ref{F4} illustrates the dependence of $v(\tau_d) \equiv |\langle \dot x (\tau_d) \rangle|$ on $\tau_d$ for high [panel (a)] and low noise intensities [panel (b)]; in both panels $\tau_c=0$. 
The curves $v(\tau_d)$ are singular at the origin: (i) The rectification speed $v(0)$ is fairly large, in close agreement with the predictions of the Fokker-Planck formalism \cite{risken} (not shown); (ii) For $\tau_d \to 0+$ the net speed $v(\tau_d)$ drops to a much lower non-zero value $v(0+) \equiv \lim_{\tau_d \to 0}|\langle \dot x (\tau_d) \rangle|$. Moreover, for $\tau_d >0$ the curves $v(\tau_d)$ exhibit a persistent tail that cannot be explained as a color effect (we recall that here $\tau_c=0$).

In the low noise regime, $Q_1 \ll 2a$ and $Q_2 \ll 1$, $v(\tau_d)$ decays exponentially with fitting law
\begin{equation}
\label{3.3}
v(\tau_d) = v(0+) e^{-a\tau_d},
\end{equation}
whereas at higher noise levels the curves $v(\tau_d)$ develop a resonant behavior. The low-noise $\tau_d$ dependence shows an obvious resemblance with the Brownian relaxation in a parabolic potential driven by a {\it linear} superposition of  $\xi_1$ and $\xi_2$, see in particular Eq. (\ref{2.3}); indeed, when sitting around the bottom of the substrate wells, the particle is subjected to two effective additive noises (harmonic or Gaussian approximation), like in Sec. II. In such approximation the exponential decay (\ref{3.3}) can be explained with the progressive decorrelation of the recycled versus the primary noise with increasing $\tau_d$. The resonant $\tau_d$ dependence of the rectification effect at higher $Q_1,Q_2$, instead, can be traced back to an optimal synchronization of the additive, $\xi_1$, and multiplicative fluctuations, $\xi_2$, which occurs only because of the {\it nonlinearity} of the substrate $V(x)$. This and related aspects of the present rectification mechanism will be discussed in Sec. IV.

On computing the characteristics (or response) function $v(\tau_d)$, we remark that the net current $\langle \dot x (\tau_d) \rangle$ obeys obvious symmetry relations that allow us to restrict the parameter range to explore: (i) $\langle \dot x (-\tau_d) \rangle=\langle \dot x (\tau_d) \rangle$; (ii) $\langle \dot x (\tau_d) \rangle \to -\langle \dot x (\tau_d) \rangle$ upon changing the relative sign of $\xi_1$ and $\xi_2$. Symmetries (i) and (ii) are easy to prove both analytically and numerically (not shown).

The peculiar $\tau_d$ dependence of the rectification function $v(\tau_d)$ is important in view of practical applications. Indeed, in many circumstances, it would be extremely difficult to recycle a control signal $\xi_2(t)$ so that $\tau_d \ll \tau_c$; stated otherwise, measuring $v(0)$ requires a certain degree of experimental sophistication. On the contrary, if we agree to work on the resonant tail of its response curve $v(\tau_d)$, a rectification device described by the SDE (\ref{3.2}) can be operated with less effort; the net ouput current is not the highest, as $v(\tau_d>0) < v(0)$, but is still appreciable and, more importantly, stable against the accidental floating of the control parameter $\tau_d$. 

In this sense the scheme represented in Eq. (\ref{3.2}) is a simple-minded attempt at implementing the operation of a Maxwell's daemon: The ideal device we set up is intended to gauge the primary random signal $\xi_1$ at the sampling time $t$ and, depending on the sign of each reading, to lower or raise the gate barriers accordingly at a later time $t+\tau_d$, i.e., open or close the trap door. The rectifying power of such a daemon is far from optimal; lacking the dexterity of Maxwell's ''gate-keeper'' \cite{daemon}, it works ''in average'' like a ''dumb'' automaton.

\begin{figure}[ht]
\includegraphics*[width=8.0cm]{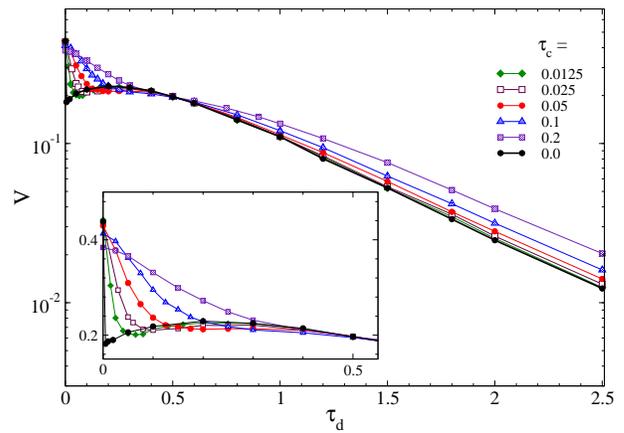}
\caption{\label{F5}  Characteristics curve $v(\tau_d)$ for the process (\ref{3.2}) with $a=1$, $D\equiv 0$, $Q_1=Q_2=1$, and different autocorrealtion times $\tau_c$.}
\end{figure}

\section{Rectification properties}

We discuss now the dependence of the response function $v(\tau_d)$ on the remaining parameters of the process (\ref{3.2}).

\begin{figure}[ht]
\includegraphics*[width=8.0cm]{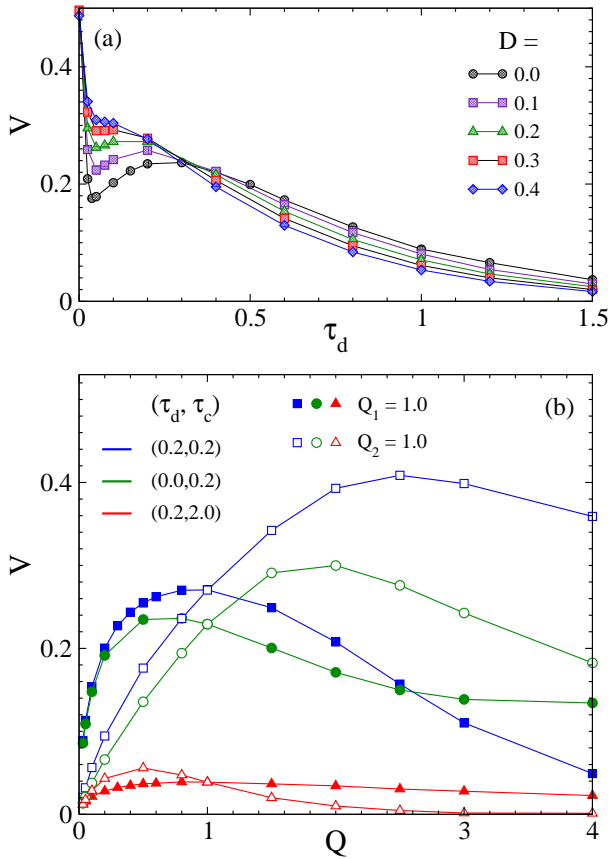}
\caption{\label{F6} (Color online) (a) Characteristics curve $v(\tau_d)$ for the process (\ref{3.2}) with $a=1$, $Q_1=Q_2=1$, $\tau_c=0$, and different $D$. (b) $v$ vs. $Q_2$ with $Q_1=1$ (empty symbols) and vs. $Q_1$ with $Q_2=1$ (filled symbols) for different values of $(\tau_d,\tau_c)$: squares $(0.2,0.2)$, circles $(0.0, 0.2)$, and triangles $(0.2,2.0)$. The remaining simulation parameters are $D\equiv 0$ and $a=1$.}
\end{figure}

(1) {\it Dependence on the noise autocorrelation time $\tau_c$.} As reminded in Ref. \cite{confinement}, a finite correlation time $\tau_c$ of the cooperating noises $\xi_1$ and $\xi_2$ enhances the asymmetry of the system response against the relative delay $\tau_d$. In particular, the discontinuity $v(0)$ versus $v(0+)$ in Fig. \ref{F4}(a) for $\tau_c=0$, is replaced by a continuous drop of $v(\tau_d)$, which approaches its resonant tail down from $v(0)$ on a scale $\tau_d \sim \tau_c$; conversely, the tails of $v(\tau_d)$ decay exponentially more slowly on increasing $\tau_c$, until their resonance peak disappears completely;

(2) {\it Dependence on the noise crosscorrelation.} So far, the recycled and primary noise have been assumed to be fully cross-correlated, that is $\xi_1$ and $\xi_2$ are both proportional to the same source $\eta$; the relative time delay $\tau_d$ determines an effective decorrelation in their action on the diffusing particle. We now switch on the thermal noise $\epsilon (t)$ in Eq. (\ref{3.2}):  The recycled noise $\xi_2$ turns out to be only partially correlated with the {\it total} additive noise $\xi_1 + \epsilon$, which drives the process with intensity $Q_1+D$. Thermal noise tends to foil the synchronized effort of $\xi_1$ and $\xi_2$; this is a circumstance of current interest in real experiments \cite{virgo}. On decreasing the multiplicative-additive noise crosscorrelation, e.g., by increasing $D$, the tails of $v(\tau_d)$ in Fig. \ref{F6} become less and less persistent, i.e., decay faster and faster; their resonance peak shifts towards lower $\tau_d$ values, until it merges into the narrow peak centered at $\tau_d=0$ \cite{grigolini};

(3) {\it Dependence on the recycled-to-primary noise intensity ratio.} The net rectification current speed $v$ exhibits a clearcut resonant dependence on both $Q_1$ and $Q_2$, separately, regardless of the time constants $\tau_c$ and $\tau_d$ -- see Fig. \ref{F6}(b). This is certainly true for $\tau_d=0$, as known from the Fokker-Planck formalism \cite{res.activation}, but such an effect is more pronounced for intermediate $\tau_c, \tau_d$, namely $0<\tau_c, \tau_d<a$. Under optimal rectification the maxima of $v$ vs. $Q$ in Fig. \ref{F6}(b) occur, as expected, for ${Q_1} _{\, \sim}^{\, <} \;1$ and ${Q_2} _{\, \sim}^{\, <} \;2a$.

\section{Conclusions}

The model (\ref{3.2}) can be regarded as the prototype of a rectifying device; inspired by biological systems \cite{ratchet}, this scheme can be employed to design and operate artificial rectifiers, for instance, of magnetic vortices and colloidal particles. The dependence of the response function $v(\tau_d)$ on the noise parameters, items (1)-(3) of this section, indicated how to optimize the net rectification current across the device.

The present investigation was based on the Langevin equation formalism -- see the analytical predictions of Sec. II and the numerical results of Secs. II-IV. Alternately, one could try to formulate the description of a stochastic process $x(t)$ driven by delayed noises in term of the probability density $P(x,t)$, namely by generalizing the Fokker-Planck (FP) equation formalism \cite{risken}. For instance, under certain restrictions, the linear SDE (\ref{2.1}) corresponds to the time-delayed FP equation
\begin{eqnarray}
\label{5.1}
&&\partial_x P(x,t)= a \partial_x xP(x,t) +(Q_1+Q_2)\partial_x^2P(x,t) \nonumber \\
&\,&\pm \sqrt{Q_1Q_2}\partial_x^2[P(x,t+\tau_d)+P(x,t-\tau_d)].
\end{eqnarray}
On assuming the asymptotic time-dependence $P(x,t) = e^{-at}P(x)$, for $t\to \infty$ the FP equation (\ref{5.1}) approaches the standard form
\begin{eqnarray}
\label{5.2}
&&\partial_x P(x,t)= a \partial_x xP(x,t) \\ \nonumber
&\,&+[(Q_1+Q_2) \pm \sqrt{Q_1Q_2}(e^{a\tau_d}+e^{-a\tau_d}] \partial_x^2P(x,t),
\end{eqnarray}
which correctly reproduces the long time behavior of Eq. (\ref{2.4}) -- but not the stationary quantity Eq. (\ref{2.3}).
A rigorous FP equation formalism for SDE with delayed noises, valid at all times, is the subject of ongoing research.

\end{document}